\documentclass[reprint,amsmath,amssymb,aps,prl,longbibliography,groupaddress]{revtex4-2}
\usepackage{graphicx}
\usepackage{dcolumn}
\usepackage{bm}
\usepackage{amsmath}
\usepackage{placeins}
\usepackage{epsfig}
\begin{document}
\title{Comment on ``Experimental Evidence for a State-Point-Dependent Density-Scaling Exponent of Liquid Dynamics"}
\author{T. C. Ransom}
\altaffiliation{American Society for Engineering Education postdoc}
\email{timothy.ransom.ctr@nrl.navy.mil}
\author{R. Casalini}
\author{D. Fragiadakis}
\author{A. P. Holt}
\altaffiliation{American Society for Engineering Education postdoc}
\author{C. M. Roland}
\email{mike.roland@nrl.navy.mil}
\affiliation{Naval Research Laboratory, Chemistry Division, Washington DC 20375-5342}
\maketitle
\begin{figure}
\includegraphics[width=\linewidth]{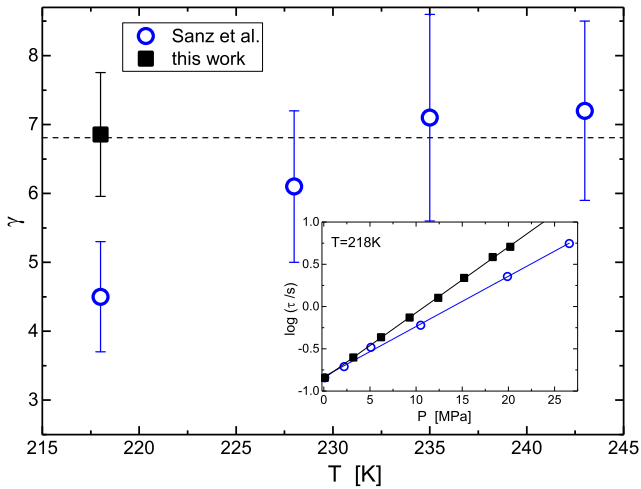}
	\caption{\label{fig:gamma}Scaling exponent for DC704 from ref.~\cite{sanz2019experimental} and the new result. Inset shows the pressure dependence of $\tau$ at 218K which in Sanz et al.~has a slope anomalously decreasing beyond 5 MPa.}
\end{figure}
\begin{figure}
\includegraphics[width=\linewidth]{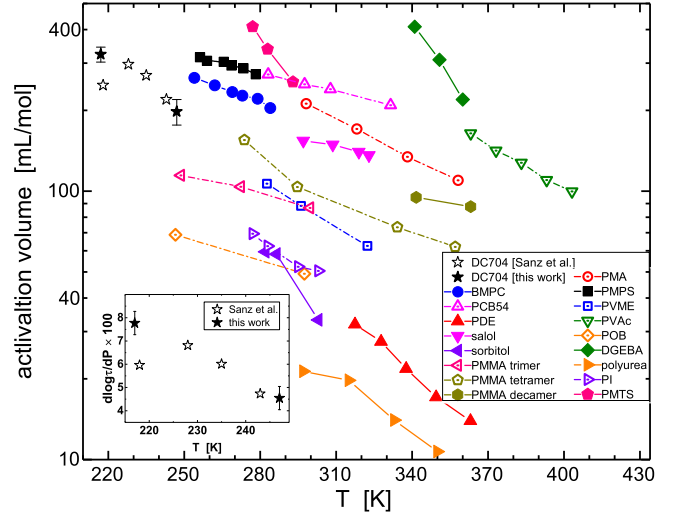}
    \caption{\label{fig:activationV}Activation volumes for 18 materials, including data for DC704 from ref.~\cite{sanz2019experimental} and herein. The inset shows $(\partial\log\tau/\partial p)_T$ for DC704, from which $\Delta V$ is obtained.}
\end{figure}

It has been established from data on more than 100 liquids and polymers that the relaxation time and other dynamic quantities superimpose when plotted versus $T\rho^{-\gamma}$, where $\gamma$ is a material constant \cite{roland2011viscoelastic,Grzybowski2018}. The known exception to this density scaling is H-bonded and other associated liquids. Deviations from an invariant $\gamma$ of about 10\% have been observed in molecular dynamic simulations for substantial density changes, \textit{ca.}~10\% \cite{schroder2009hidden}; however, experimentally, density scaling has been verified for pressures as high as 10 GPa in diamond anvil measurements \cite{ransom2017vitrification,ransom2017glass,abramson2014viscosity}. Recently Sanz et al.~\cite{sanz2019experimental} reported that the scaling exponent $\gamma$ for two simple liquids were state-point dependent, with data presented for one of these materials, tetramethyl-tetraphenyl-trisiloxane (DC704). Their reported $\gamma$ is shown in Figure~\ref{fig:gamma}, where deviation from a constant $\gamma$ is seen for one point at the lowest temperature, 218K. In ref.~\cite{sanz2019experimental} $\gamma$ were calculated using the formula
\begin{equation}\label{Eq:gamma}
    \gamma=-\frac{K_T(\partial\log\tau/\partial p)_T}{T(\partial\log\tau/\partial T)_p+\alpha_PTK_T(\partial\log\tau/\partial p)_T}
\end{equation}
in which $K_T$ is the isothermal bulk modulus and $\alpha_P$ is the isobaric thermal expansion coefficient. The error in Fig.~\ref{fig:gamma} comes from the quantity $(\partial\log\tau/\partial p)_T$ at $T=218$K, which Sanz et al. reported as decreasing with increasing $p$.
This is an unphysical result; after an initial linear dependence, relaxation times increase more strongly with increasing pressure. To show that the result is at odds with available data, in Figure~\ref{fig:activationV} are plotted activation volumes, $\Delta V=RT(d\ln\tau/dp)_T$, for 18 substances. Excepting the result for DC704 from ref.~\cite{sanz2019experimental}, all show a decrease in $\Delta V$ as $T$ increases. \\
\indent The underestimate of the pressure coefficient of $\tau$ at T=218K causes $\gamma$ calculated from eq.~(\ref{Eq:gamma}) to be underestimated at this temperature. We re-measured the pressure coefficient of $\tau$ for DC704, and as seen in Fig.~\ref{fig:gamma}, there is no decrease in $(\partial\log\tau/\partial p)_T$  at higher $p$. Using the new data $\gamma$ is recalculated (eq.~\ref{Eq:gamma}), with the new result included in Fig.~\ref{fig:gamma}. The scaling exponent for DC704 is indeed invariant within uncertainty over the studied range of $T$ and $p$.\\
\indent In summary, the substantial variation of the scaling exponent (44\% change in $\gamma$ for a 2\% change in density) reported for DC704 in ref.~\cite{sanz2019experimental} is a result of an erroneous measurement of the pressure dependence of $\tau$ at low temperature. The correct value of $(\partial\log\tau/\partial p)_T$  yields a $\gamma$ that is state-point independent within uncertainty, consistent both with previous publications on this particular liquid \cite{gundermann2011predicting,casalini2011density} and with the prodigious amount of existing data on simple liquids \cite{roland2011viscoelastic,Grzybowski2018}. While variation of $\gamma$ with $T$ and $p$ is known from simulations, the evidence to support this in real materials is currently lacking.\\
\indent TR and AH acknowledge an ASEE postdoctoral fellowship. We thank A. Sanz for kindly providing their data for DC704. This work was supported by the Office of Naval Research. 
\providecommand{\noopsort}[1]{}\providecommand{\singleletter}[1]{#1}%

\providecommand{\noopsort}[1]{}\providecommand{\singleletter}[1]{#1}%
\begin{thebibliography}{9}%
\makeatletter
\providecommand \@ifxundefined [1]{%
 \@ifx{#1\undefined}
}%
\providecommand \@ifnum [1]{%
 \ifnum #1\expandafter \@firstoftwo
 \else \expandafter \@secondoftwo
 \fi
}%
\providecommand \@ifx [1]{%
 \ifx #1\expandafter \@firstoftwo
 \else \expandafter \@secondoftwo
 \fi
}%
\providecommand \natexlab [1]{#1}%
\providecommand \enquote  [1]{``#1''}%
\providecommand \bibnamefont  [1]{#1}%
\providecommand \bibfnamefont [1]{#1}%
\providecommand \citenamefont [1]{#1}%
\providecommand \href@noop [0]{\@secondoftwo}%
\providecommand \href [0]{\begingroup \@sanitize@url \@href}%
\providecommand \@href[1]{\@@startlink{#1}\@@href}%
\providecommand \@@href[1]{\endgroup#1\@@endlink}%
\providecommand \@sanitize@url [0]{\catcode `\\12\catcode `\$12\catcode
  `\&12\catcode `\#12\catcode `\^12\catcode `\_12\catcode `\%12\relax}%
\providecommand \@@startlink[1]{}%
\providecommand \@@endlink[0]{}%
\providecommand \url  [0]{\begingroup\@sanitize@url \@url }%
\providecommand \@url [1]{\endgroup\@href {#1}{\urlprefix }}%
\providecommand \urlprefix  [0]{URL }%
\providecommand \Eprint [0]{\href }%
\providecommand \doibase [0]{https://doi.org/}%
\providecommand \selectlanguage [0]{\@gobble}%
\providecommand \bibinfo  [0]{\@secondoftwo}%
\providecommand \bibfield  [0]{\@secondoftwo}%
\providecommand \translation [1]{[#1]}%
\providecommand \BibitemOpen [0]{}%
\providecommand \bibitemStop [0]{}%
\providecommand \bibitemNoStop [0]{.\EOS\space}%
\providecommand \EOS [0]{\spacefactor3000\relax}%
\providecommand \BibitemShut  [1]{\csname bibitem#1\endcsname}%
\let\auto@bib@innerbib\@empty
%</preamble>
\bibitem [{\citenamefont {Roland}(2011)}]{roland2011viscoelastic}%
  \BibitemOpen
  \bibfield  {author} {\bibinfo {author} {\bibfnamefont {C.~M.}\ \bibnamefont
  {Roland}},\ }\href@noop {} {\emph {\bibinfo {title} {Viscoelastic behavior of
  rubbery materials}}}\ (\bibinfo  {publisher} {Oxford University Press},\
  \bibinfo {year} {2011})\BibitemShut {NoStop}%
\bibitem [{\citenamefont {Grzybowski}\ and\ \citenamefont
  {Paluch}(2018)}]{Grzybowski2018}%
  \BibitemOpen
  \bibfield  {author} {\bibinfo {author} {\bibfnamefont {A.}~\bibnamefont
  {Grzybowski}}\ and\ \bibinfo {author} {\bibfnamefont {M.}~\bibnamefont
  {Paluch}},\ }in\ \href@noop {} {\emph {\bibinfo {booktitle} {The Scaling of
  Relaxation Processes}}},\ \bibinfo {editor} {edited by\ \bibinfo {editor}
  {\bibfnamefont {F.}~\bibnamefont {Kremer}}\ and\ \bibinfo {editor}
  {\bibfnamefont {A.}~\bibnamefont {Loidl}}}\ (\bibinfo  {publisher}
  {Springer},\ \bibinfo {year} {2018})\BibitemShut {NoStop}%
\bibitem [{\citenamefont {Schr{\o}der}\ \emph {et~al.}(2009)\citenamefont
  {Schr{\o}der}, \citenamefont {Pedersen}, \citenamefont {Bailey},
  \citenamefont {Toxvaerd},\ and\ \citenamefont {Dyre}}]{schroder2009hidden}%
  \BibitemOpen
  \bibfield  {author} {\bibinfo {author} {\bibfnamefont {T.~B.}\ \bibnamefont
  {Schr{\o}der}}, \bibinfo {author} {\bibfnamefont {U.~R.}\ \bibnamefont
  {Pedersen}}, \bibinfo {author} {\bibfnamefont {N.~P.}\ \bibnamefont
  {Bailey}}, \bibinfo {author} {\bibfnamefont {S.}~\bibnamefont {Toxvaerd}},\
  and\ \bibinfo {author} {\bibfnamefont {J.~C.}\ \bibnamefont {Dyre}},\
  }\href@noop {} {\bibfield  {journal} {\bibinfo  {journal} {Phys. Rev. E}\
  }\textbf {\bibinfo {volume} {80}},\ \bibinfo {pages} {041502} (\bibinfo
  {year} {2009})}\BibitemShut {NoStop}%
\bibitem [{\citenamefont {Ransom}\ \emph {et~al.}(2017)\citenamefont {Ransom},
  \citenamefont {Ahart}, \citenamefont {Hemley},\ and\ \citenamefont
  {Roland}}]{ransom2017vitrification}%
  \BibitemOpen
  \bibfield  {author} {\bibinfo {author} {\bibfnamefont {T.~C.}\ \bibnamefont
  {Ransom}}, \bibinfo {author} {\bibfnamefont {M.}~\bibnamefont {Ahart}},
  \bibinfo {author} {\bibfnamefont {R.~J.}\ \bibnamefont {Hemley}},\ and\
  \bibinfo {author} {\bibfnamefont {C.~M.}\ \bibnamefont {Roland}},\
  }\href@noop {} {\bibfield  {journal} {\bibinfo  {journal} {Macromolecules}\
  }\textbf {\bibinfo {volume} {50}},\ \bibinfo {pages} {8274} (\bibinfo {year}
  {2017})}\BibitemShut {NoStop}%
\bibitem [{\citenamefont {Ransom}\ and\ \citenamefont
  {Oliver}(2017)}]{ransom2017glass}%
  \BibitemOpen
  \bibfield  {author} {\bibinfo {author} {\bibfnamefont {T.~C.}\ \bibnamefont
  {Ransom}}\ and\ \bibinfo {author} {\bibfnamefont {W.~F.}\ \bibnamefont
  {Oliver}},\ }\href@noop {} {\bibfield  {journal} {\bibinfo  {journal} {Phys.
  Rev. Lett.}\ }\textbf {\bibinfo {volume} {119}},\ \bibinfo {pages} {025702}
  (\bibinfo {year} {2017})}\BibitemShut {NoStop}%
\bibitem [{\citenamefont {Abramson}(2014)}]{abramson2014viscosity}%
  \BibitemOpen
  \bibfield  {author} {\bibinfo {author} {\bibfnamefont {E.~H.}\ \bibnamefont
  {Abramson}},\ }\href@noop {} {\bibfield  {journal} {\bibinfo  {journal} {J.
  Phys. Chem. B}\ }\textbf {\bibinfo {volume} {118}},\ \bibinfo {pages} {11792}
  (\bibinfo {year} {2014})}\BibitemShut {NoStop}%
\bibitem [{\citenamefont {Sanz}\ \emph {et~al.}(2019)\citenamefont {Sanz},
  \citenamefont {Hecksher}, \citenamefont {Hansen}, \citenamefont {Dyre},
  \citenamefont {Niss},\ and\ \citenamefont {Pedersen}}]{sanz2019experimental}%
  \BibitemOpen
  \bibfield  {author} {\bibinfo {author} {\bibfnamefont {A.}~\bibnamefont
  {Sanz}}, \bibinfo {author} {\bibfnamefont {T.}~\bibnamefont {Hecksher}},
  \bibinfo {author} {\bibfnamefont {H.~W.}\ \bibnamefont {Hansen}}, \bibinfo
  {author} {\bibfnamefont {J.~C.}\ \bibnamefont {Dyre}}, \bibinfo {author}
  {\bibfnamefont {K.}~\bibnamefont {Niss}},\ and\ \bibinfo {author}
  {\bibfnamefont {U.~R.}\ \bibnamefont {Pedersen}},\ }\href@noop {} {\bibfield
  {journal} {\bibinfo  {journal} {Phys. Rev. Lett.}\ }\textbf {\bibinfo
  {volume} {122}},\ \bibinfo {pages} {055501} (\bibinfo {year}
  {2019})}\BibitemShut {NoStop}%
\bibitem [{\citenamefont {Gundermann}\ \emph {et~al.}(2011)\citenamefont
  {Gundermann}, \citenamefont {Pedersen}, \citenamefont {Hecksher},
  \citenamefont {Bailey}, \citenamefont {Jakobsen}, \citenamefont
  {Christensen}, \citenamefont {Olsen}, \citenamefont {Schr{\o}der},
  \citenamefont {Fragiadakis}, \citenamefont {Casalini} \emph
  {et~al.}}]{gundermann2011predicting}%
  \BibitemOpen
  \bibfield  {author} {\bibinfo {author} {\bibfnamefont {D.}~\bibnamefont
  {Gundermann}}, \bibinfo {author} {\bibfnamefont {U.~R.}\ \bibnamefont
  {Pedersen}}, \bibinfo {author} {\bibfnamefont {T.}~\bibnamefont {Hecksher}},
  \bibinfo {author} {\bibfnamefont {N.~P.}\ \bibnamefont {Bailey}}, \bibinfo
  {author} {\bibfnamefont {B.}~\bibnamefont {Jakobsen}}, \bibinfo {author}
  {\bibfnamefont {T.}~\bibnamefont {Christensen}}, \bibinfo {author}
  {\bibfnamefont {N.~B.}\ \bibnamefont {Olsen}}, \bibinfo {author}
  {\bibfnamefont {T.~B.}\ \bibnamefont {Schr{\o}der}}, \bibinfo {author}
  {\bibfnamefont {D.}~\bibnamefont {Fragiadakis}}, \bibinfo {author}
  {\bibfnamefont {R.}~\bibnamefont {Casalini}}, \emph {et~al.},\ }\href@noop {}
  {\bibfield  {journal} {\bibinfo  {journal} {Nature Physics}\ }\textbf
  {\bibinfo {volume} {7}},\ \bibinfo {pages} {816} (\bibinfo {year}
  {2011})}\BibitemShut {NoStop}%
\bibitem [{\citenamefont {Casalini}\ \emph {et~al.}(2011)\citenamefont
  {Casalini}, \citenamefont {Gamache},\ and\ \citenamefont
  {Roland}}]{casalini2011density}%
  \BibitemOpen
  \bibfield  {author} {\bibinfo {author} {\bibfnamefont {R.}~\bibnamefont
  {Casalini}}, \bibinfo {author} {\bibfnamefont {R.}~\bibnamefont {Gamache}},\
  and\ \bibinfo {author} {\bibfnamefont {C.}~\bibnamefont {Roland}},\
  }\href@noop {} {\bibfield  {journal} {\bibinfo  {journal} {J. Chem. Phys.}\
  }\textbf {\bibinfo {volume} {135}},\ \bibinfo {pages} {224501} (\bibinfo
  {year} {2011})}\BibitemShut {NoStop}%
\end{thebibliography}%
\end{document}